\documentclass[twocolumn,preprintnumbers,aps,pra,superscriptaddress,amsmath,amssymb]{revtex4-1}

\usepackage{graphicx}
\usepackage{dcolumn}
\usepackage{bm}
\usepackage{bbm}
\usepackage{hyperref}
\usepackage{color}
\usepackage{multirow}
\usepackage{physics}

\hypersetup{
	colorlinks,
	citecolor=black,
	filecolor=black,
	linkcolor=black,
	urlcolor=black
}

\begin{document}

\title{Imaging graphene field-effect transistors on diamond using nitrogen-vacancy microscopy}

\author{Scott E. Lillie}
\affiliation{Centre for Quantum Computation and Communication Technology, School of Physics, The University of Melbourne, VIC 3010, Australia}
\affiliation{School of Physics, The University of Melbourne, VIC 3010, Australia}

\author{Nikolai Dontschuk}
\email{dontschuk.n@unimelb.edu.au}
\affiliation{Centre for Quantum Computation and Communication Technology, School of Physics, The University of Melbourne, VIC 3010, Australia}
\affiliation{School of Physics, The University of Melbourne, VIC 3010, Australia}

\author{David A. Broadway}
\affiliation{Centre for Quantum Computation and Communication Technology, School of Physics, The University of Melbourne, VIC 3010, Australia}
\affiliation{School of Physics, The University of Melbourne, VIC 3010, Australia}

\author{Daniel L. Creedon}
\affiliation{School of Physics, The University of Melbourne, VIC 3010, Australia}

\author{Lloyd C. L. Hollenberg}
\affiliation{Centre for Quantum Computation and Communication Technology, School of Physics, The University of Melbourne, VIC 3010, Australia}
\affiliation{School of Physics, The University of Melbourne, VIC 3010, Australia}

\author{Jean-Philippe Tetienne}
\email{jtetienne@unimelb.edu.au}
\affiliation{School of Physics, The University of Melbourne, VIC 3010, Australia}

\date{\today}

\begin{abstract}

The application of imaging techniques based on ensembles of nitrogen-vacancy (NV) sensors in diamond to characterise electrical devices has been proposed, but the compatibility of NV sensing with operational gated devices remains largely unexplored. Here we fabricate graphene field-effect transistors (GFETs) directly on the diamond surface and characterise them via NV microscopy. The current density within the gated graphene is reconstructed from NV magnetometry under both mostly p- and n-type doping, but the exact doping level is found to be affected by the measurements. Additionally, we observe a surprisingly large modulation of the electric field at the diamond surface under an applied gate potential, seen in NV photoluminescence and NV electrometry measurements, suggesting a complex electrostatic response of the oxide-graphene-diamond structure. Possible solutions to mitigate these effects are discussed.

\end{abstract}

\maketitle

\section{Introduction}

Sensing techniques using nitrogen-vacancy (NV) centres in diamond \cite{Doherty2013} provide a convenient platform by which condensed matter systems can be interrogated \cite{Casola2018}. The local sensitivity of NV centres to both magnetic and electric fields, in concert with their flexible experimental conditions \cite{Rondin2014}, permit both sensing and imaging of a range of nano- to meso-scale phenomena. An application of these techniques is the characterisation of electrical devices and related materials. Magnetic noise spectroscopy has been used to probe and map the local conductivity of thin metallic films \cite{Kolkowitz2015,Ariyaratne2018}, lending insight to the motion of carriers within the material, and also to identify noise sources from sparse metallic depositions \cite{Lillie2018}. In low dimensional systems, magnetic noise spectroscopy should provide insight into local electronic correlations \cite{Agawal2017,Rodriguez-Nieva2018a}, whereas static magnetic field mapping has been used to map charge transport in carbon nano-tubes using scanning single-NV centres \cite{Chang2017}, and monolayer graphene ribbons via wide-field imaging experiments \cite{Tetienne2017}. Extensions to the latter scenario should be able to measure signatures of viscous electron flow in graphene and similar systems \cite{Bandurin2016,Guerrero-Becerra2019,Ku2019}, and other mesoscopic effects such as electron-hole puddling \cite{Martin2008}, and gate controlled steering of carriers \cite{Williams2011}.

Wide-field imaging of electrical devices using NV ensembles generally requires the devices to be fabricated directly on the NV-diamond substrate, while many interesting transport phenomena require precise control over the doping in the conductive channel. This can be achieved in a field-effect transistor (FET) device, employing either a top gate \cite{Kim2008,Andersen2019,Ku2019}, an in-plane gate \cite{Hahnlein2012,Hauf2014}, or an electrolytic gate \cite{Ohno2009}. Recently, graphene based devices have been fabricated successfully on NV diamond substrates \cite{Tetienne2017,Andersen2019,Ku2019}, but the compatibility of such structures with wide-field NV microscopy remains largely unexplored, with questions of whether the operating conditions required for NV microscopy may affect the operation and integrity of the FET, or whether the fabrication/operation of the FET may affect the ability to perform NV sensing.

In this work, we fabricate a number of top-gated graphene field-effect transistors (GFETs) on an NV-diamond substrate, and characterise device phenomena via wide-field imaging of the near-surface NV ensemble. Current is injected into the graphene ribbons, $I_{SD}$, by application of a source-drain potential, $V_{SD}$, and the doping of the ribbon is tunable via a top gate potential, $V_G$, allowing charge transport to be probed in different doping regimes, and for the effect of the gate to be studied [Fig. \ref{Fig1}(a)]. Firstly, the devices are characterised by electrical measurements and the influence of the laser, used to excite the NV-layer, is assessed. Secondly, the current density within the graphene ribbon is reconstructed under p- and n-type doping by measuring the associated \O rsted field via optical readout of the NV electron-spin resonances. Thirdly, an effect is observed by which the NV-layer photoluminescence (PL) is modulated by the applied gate potential in regions proximal to the gated device, but extending up to $20$\,$\mu$m away from the graphene ribbon. Direct measurement of the electric field by the NV ensemble electron-spin resonances demonstrate that this effect is due to an enhanced electric field surrounding the graphene ribbon which diminishes the NV$^-$/NV$^0$ charge state ratio. Finally, we discuss possible solutions to overcome the challenges identified in this study, to facilitate further investigation of transport phenomena in graphene and other two-dimensional materials.

\section{Fabrication}

\begin{figure}[t]
	\begin{center}
		\includegraphics[width=1.0\columnwidth]{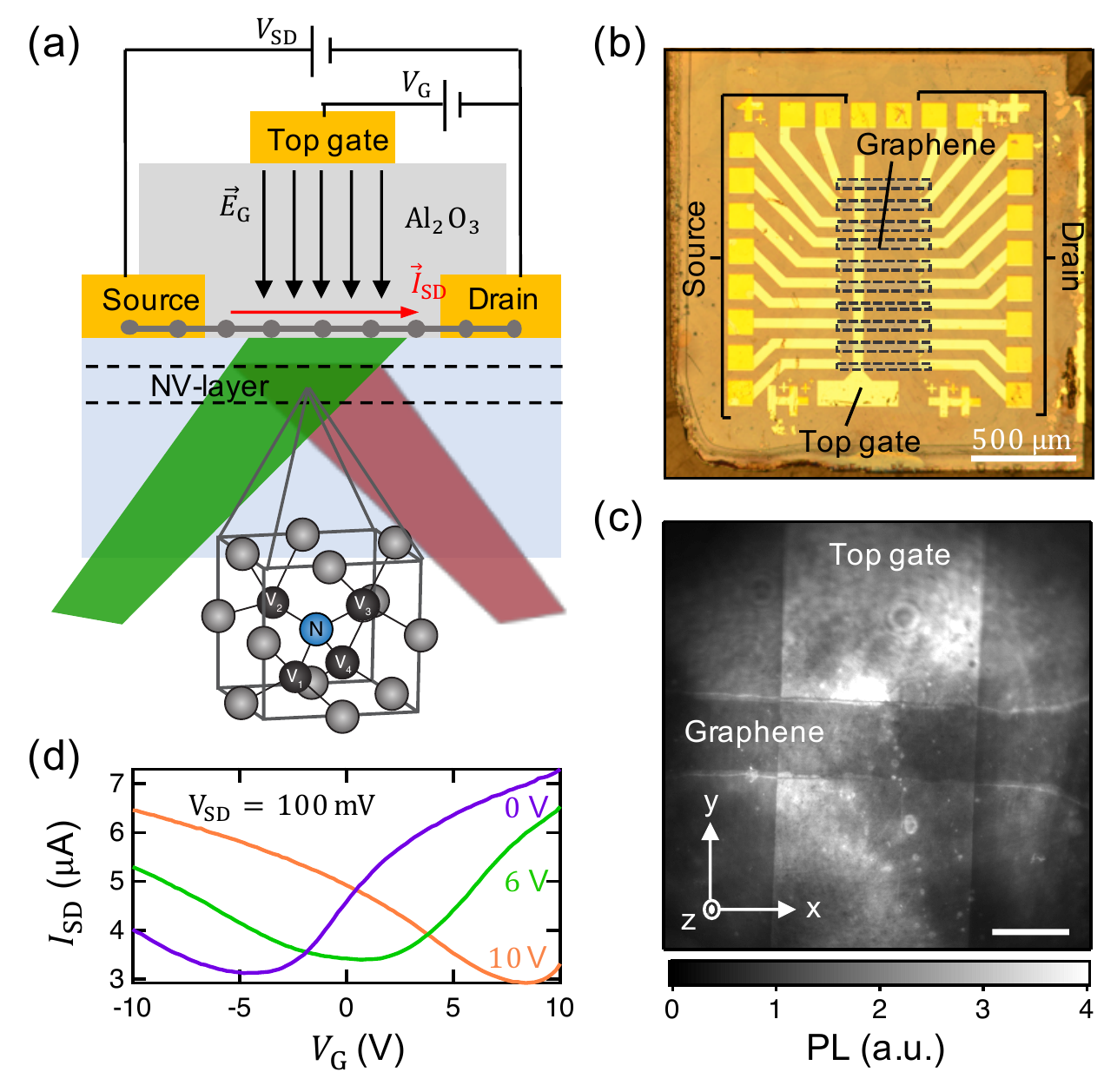}
		\caption{\textbf{Imaging GFETs with NV-diamond:} (a) Top-gated graphene field-effect transistor (GFET) fabricated on an NV-diamond substrate. Current is injected into the graphene ribbon, $I_{SD}$, by application of a source-drain potential, $V_{SD}$, while the graphene is doped electrically by application of a top gate potential, $V_{G}$, which is referenced to the drain.  An ensemble of NV centres is formed $10-20$\,nm below the diamond surface, and exist along all four possible crystallographic orientations, represented by the four possible vacancy positions (V$_{1,2,3,4}$) relative to a given N. Readout of the NV-layer is achieved optically, using a $532$\,nm laser for excitation (green) and the NV photoluminescence (red) is collected on an sCMOS camera. (b) Photograph of a series of devices fabricated on an NV-diamond substrate. The scale bar is $500$\,$\mu$m. (c) Photoluminescence (PL) image of a single GFET. The coordinate system is defined and the scale bar is $20$\,$\mu$m. (d) $I_{SD}$ versus $V_{G}$ curves of a single device measured in the dark after being left to equilibrate at $V_G = 0$\,V, $6$\,V, and $10$\,V for the purple, green, and orange curves respectively, under CW laser illumination, which shifts the conductivity minimum via a photo-doping effect (appendix \ref{eleccharac}).}
		\label{Fig1}
	\end{center}
\end{figure}

Graphene field-effect transistors (GFETs) were fabricated on NV-diamond substrates by using a standard wet chemical method \cite{Liang2011} to transfer monolayer graphene to the substrate from commercially available CVD polycrystalline graphene on Cu foil. The transferred graphene was selectively etched into $20$\,$\mu$m or $50$\,$\mu$m wide ribbons in an oxygen plasma with a photolithographic resist mask. Multiple photolithography steps were used to create the Cr/Au $5$/$70$\,nm source and drain contacts, wire bonding pads and the top gate contact [Fig. \ref{Fig1}(b)]. Atomic layer deposition (ALD) with modified precursor pulses was used to grow an $80$\,nm Al$_2$O$_3$ dielectric layer directly on the graphene ribbons without the use of a nucleation layer \cite{Aria2016} (appendix \ref{FETfab}). The diamond substrates used in these experiments feature an NV-layer formed by ion implantation $10-20$\,nm below the surface (appendix \ref{sample}). 

\section{Gating effect and laser}

After fabrication, the devices were imaged via the NV-layer PL, using a $532$\,nm continuous wave (CW) laser to excite the NV ensemble, and collecting the PL on an sCMOS camera, filtered around the NV$^-$ phonon side band at $690$\,nm [Fig. \ref{Fig1}(c)]. Direct visualisation of the graphene is made possible by a F\"orster resonant energy transfer (FRET) interaction between the graphene ribbon and the near-surface NV-layer, which quenches the PL \cite{Tisler2013}. The top gate appears brighter due to enhanced illumination at the NV-layer under the gate, due to a standing wave formed with the reflected light \cite{Tetienne2019}.

Electrical characterisation of the devices found source-drain resistances varying between $7$\,k$\Omega$ and $12$\,k$\Omega$ for the $50$\,$\mu$m wide devices ($100$\,$\mu$m between source to drain contacts), and between $16$\,k$\Omega$ and $91$\,k$\Omega$ for the $20$\,$\mu$m wide devices ($400$\,$\mu$m between source to drain contacts), and good Ohmic behaviour (appendix \ref{eleccharac}). Applying a small source drain potential, $V_{SD} = 100$\,mV, and measuring the source-drain current, $I_{SD}$, while sweeping the gate potential, $V_G$, a conductivity minimum was found for most devices (appendix \ref{eleccharac}). The conductivity minima, which indicate the charge neutrality point, were found to shift significantly depending on the measurement conditions and the history of the device, specifically, the applied gate potential and illumination conditions prior to the measurement [Fig. \ref{Fig1}(d)]. We attribute this to a photon assisted charge transfer between the graphene and the oxide (appendix \ref{eleccharac}), similar to the optical-doping seen with other gate dielectrics and substrates \cite{Kim2013,Tiberj2013,Ju2014a}. In addition to the photo-doping effect, we observe hysteresis in our $I_{SD}$ versus $V_G$ measurements, even when measuring in the dark (appendix \ref{eleccharac}). The hysteresis is likely due to screening of the electric field associated with the gate potential by an accumulation of trapped charge at both the graphene-oxide interface and within the oxide bulk. This has been demonstrated to cause similar hysteresis in graphene devices on SiO$_2$ \cite{Lee2011,Mao2016}, and may also be associated with the defect density within the graphene itself \cite{KrishnaBharadwaj2016}. The presence of the laser is likely to exacerbate this situation by creating additional trapped mobile charges within the oxide \cite{Ha2014}. The photo-doping induced by the laser is reproducible and sufficiently stable under fixed illumination and gate potentials to maintain the selected majority carrier type (appendix \ref{eleccharac}) over time frames compatible with current density mapping.

\begin{figure*}[t]
	\begin{center}
		\includegraphics[width=1.0\textwidth]{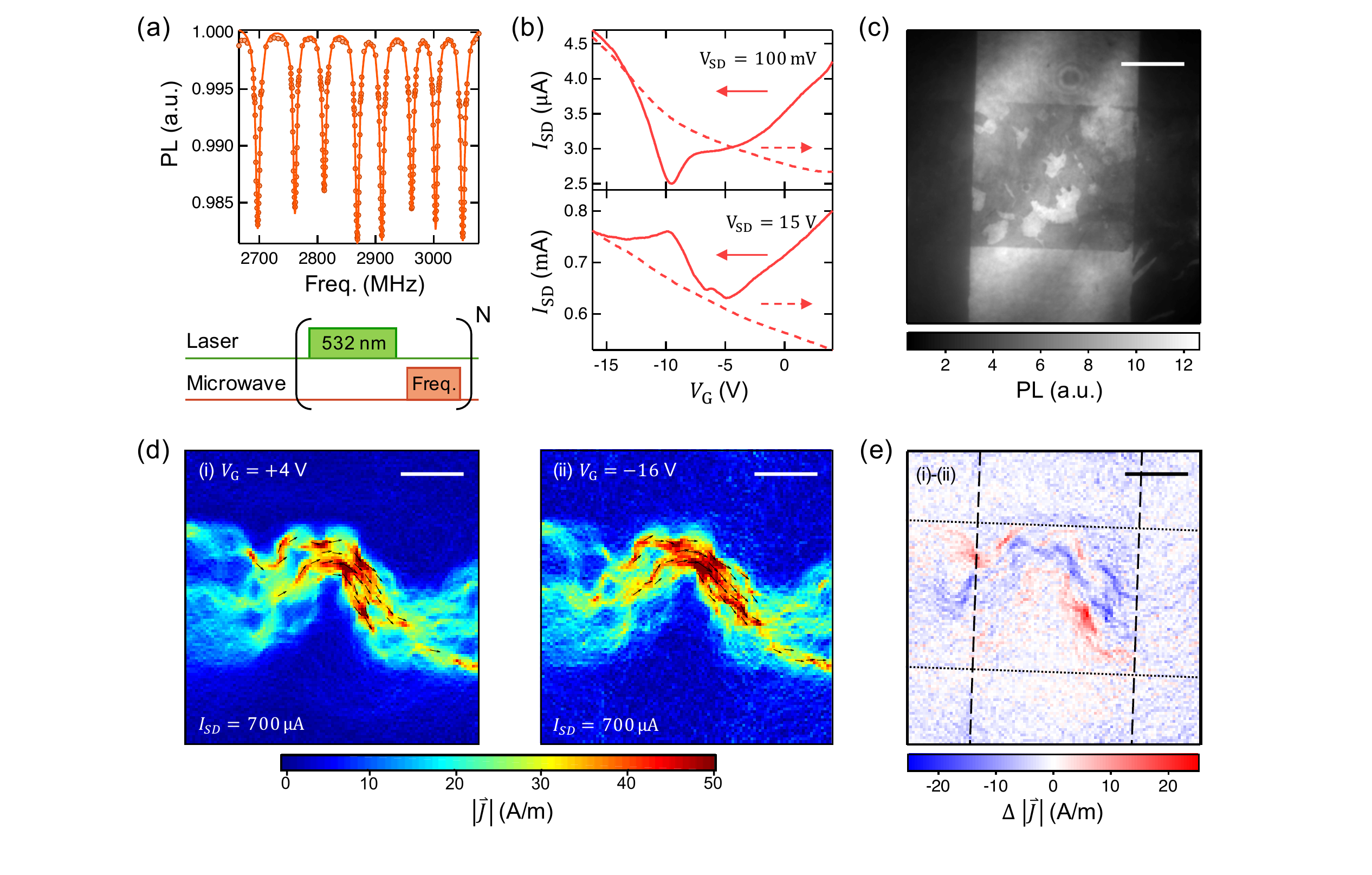}
		\caption{\textbf{Current density mapping under different doping regimes:} (a) ODMR spectrum integrated across the field of view of a single GFET device (above). The ODMR pulse sequence comprises a single laser pulse to initialise and readout the NV-layer, and a microwave pulse to drive NV spin transitions resonant with the driving frequency (below). An sCMOS camera acquires florescence over many repetitions of the pulse sequence for a given microwave frequency. (b) $I_{SD}$ versus $V_G$ measurement of the graphene device prior to current density mapping at low $V_{SD}$ (upper) and high $V_{SD}$ (lower), representative of that used during acquisition to give $I_{SD} = 700$\,$\mu$A. The device current was left to equilibrate at the initial point of each sweep under laser illumination before the gate potential was swept. Arrows denote the sweep direction. (c) PL image of the GFET device to be mapped, showing a number of tears in the graphene due to the transfer process. (d) Vector current density maps reconstructed from the \O rsted field as measured by ODMR for the GFET under n-type doping, (i) $V_G = +4$\,V, and p-type doping, (ii) $V_G = -16$\,V, each with a constant injected DC current, $I_{SD} = 700 $\,$\mu$A. Vector arrows are shown for pixels with a current density $>$\,$30$\,A/m . (e) Difference in norm current density map between the current densities at (i) $V_G = +4$\,V and (ii) $V_G = -16$\,V. The dashed (dotted) lines show the edges of the gate (graphene). All scale bars are $20$\,$\mu$m.}
		\label{Fig2}
	\end{center}
\end{figure*}

\section{Transport mapping under doping}
 
Previously, NV ensemble measurements have been used to reconstruct current densities within graphene ribbons from the associated \O rsted field as measured by optically detected magnetic resonance (ODMR) \cite{Tetienne2017}. Here we apply the same methodology, with the aim of producing current density maps in both the p-type and the n-type doped regimes. The ODMR measurement was performed with a background magnetic field oriented such that the two electron-spin resonance (ESR) transitions of each of the four NV orientations are individually resolvable [Fig. \ref{Fig2}(a)]. Each transition frequency is identified by a reduction in PL as the NV spin-state is driven from the pumped bright state, $\ket{0}$, to the less fluorescent states, $\ket{\pm1}$. The magnetic field vector at each pixel is determined by fitting the ESR frequencies, which are shifted by the graphene \O rsted field via the Zeeman effect, and determining the corresponding field projection along each of the four NV-family axes. Subtracting the background field due to the biasing permanent magnet, the measured \O rsted field is extracted, from which the current density within the graphene ribbon is reconstructed by inverting the Biot-Savart law in Fourier space (appendix \ref{currentdensity}). We note that in these devices, unlike in Ref. \cite{Tetienne2019}, no apparent current leakage into the diamond was observed (appendix \ref{currentdensity}).

Prior to mapping current densities, the device was characterised electrically to identify n- and p-type doping regimes under the conditions used for current density mapping, which require a comparatively high source-drain potential. $I_{SD}$ was measured over a $V_G$ range of $-16$\,V to $4$\,V, with $V_{SD} = 100$\,mV [Fig. \ref{Fig2}(b) upper] and $V_{SD} = 15$\,V [Fig. \ref{Fig2}(b) lower], both under CW laser illumination. Prior to each $V_G$ sweep, the device resistance was left to equilibrate at the initial gate potential under laser illumination, such that the sweep is representative of the device after photo-doping and inter-facial charge accumulation. For the decreasing gate potential sweeps (solid curves), conductivity minima are observed at $V_G=-9$\,V for the smaller $V_{SD}$, and $V_G=-5$\,V for the larger $V_{SD}$, which also shows a different shape in the transport curve. This is likely due to $V_G$ being referenced to the drain contact, and $V_G$ being of comparable magnitude to $V_{SD}$ in this scenario. For the increasing $V_G$ sweeps (dashed curves), we observe conductivity minima at $V_G\approx4$\,V and $V_G>4$\,V for the lower and higher $V_{SD}$ scenarios respectively, where the end of range is set to mitigate leakage current through the oxide. Accounting for the gate potential dependent photo-doping, we conclude that fixed gate potentials of $V_G = 4$\,V and $V_G = -16$\,V give n- and p-type doping of the graphene ribbon respectively, which should be maintained under subsequent laser pulsing (appendix \ref{eleccharac}), and hence move to acquire current density maps under each of these conditions.

A PL image of the mapped device shows the $50$\,$\mu$m wide graphene ribbon with a number of tears across the gated region, which arise during transfer and fabrication [Fig. \ref{Fig2}(c)]. Throughout imaging we maintain a constant total injected current of $700$\,$\mu$A, giving a $V_{SD}$ that varies around $15$\,V. The reconstructed current density maps of the GFET under n-type [Fig. \ref{Fig2}(d)(i)] and p-type [Fig. \ref{Fig2}(d)(ii)] doping show broadly similar features, where the current density is increased under the gated area as carriers are restricted to narrow passages due to the tears in the graphene. Taking a subtraction of the norm current density at each pixel between the two doping conditions shows clear differences in the current path [Fig. \ref{Fig2}(e)]. Some of these variations are associated with degradation of the graphene channel throughout the measurement at $V_G = -16$\,V, however, some are uncorrelated with changes to the graphene visible at our imaging resolution (appendix \ref{currentdensity}). Degradation of the devices throughout measurement precludes further investigation into the origin of these differences, however, in principle such effects could arise from inhomogeneous doping of the graphene channel, gate controlled steering of carriers \cite{Williams2011}, gate induced changes to the density of charged impurities and defects that have an asymmetric scattering cross section for electrons and holes \cite{Adam2007,Hwang2007,Wehling2010,Silvestre2013,Bai2015}.

\section{PL switching  effect}

While varying the gate potential in the previous measurement, we observed that the total PL of the NV ensemble varied significantly with the applied gate potential. To investigate this effect, a measurement was performed in which the PL was accumulated under CW laser illumination as the gate potential was varied. A settling time was introduced between setting the gate potential and starting the PL accumulation, which itself was integrated over a number of camera cycles to average out fluctuations in the illuminating laser intensity. A small source-drain potential, $V_{SD} = 100$\,mV, was applied to track the device conductivity throughout the measurement.

\begin{figure}[h!]
	\begin{center}
		\includegraphics[width=1.0\columnwidth]{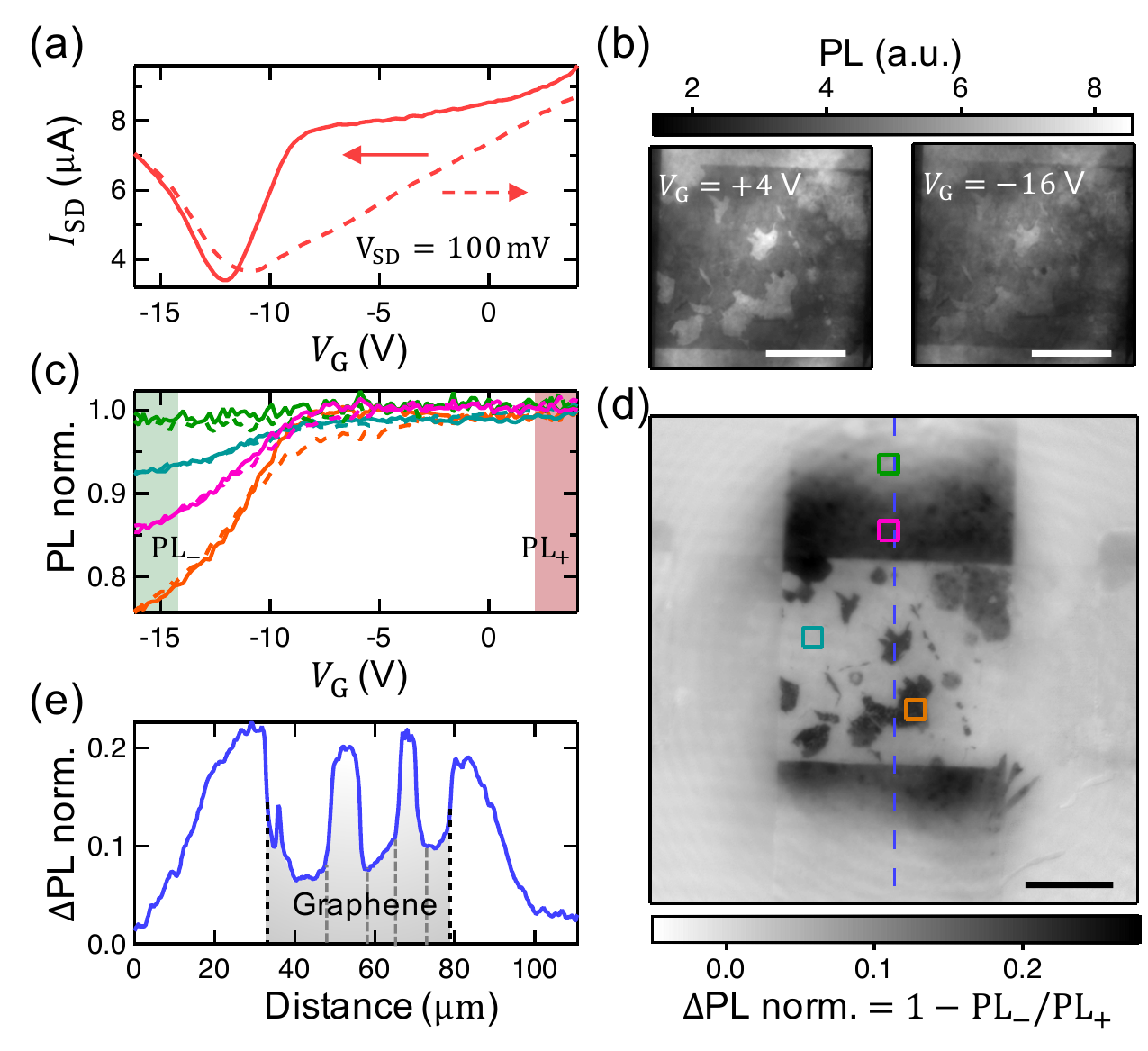}
		\caption{\textbf{Gate potential dependent PL:} (a) $I_{SD}$ versus $V_G$ measurement made in conjunction with accompanying PL measurement. The sweep moves from high to low gate potential (solid) and then reverses (dashed). (b) PL images of the gated region of a GFET at $V_G = 4$\,V and $V_G = -16$\,V showing quenched PL in the bare diamond proximal to the graphene at the lower gate potential. (c) Area normalised PL curves from four regions surrounding the GFET, specifically, a tear in the gated graphene region (orange), pristine gated graphene (blue), and two regions under the gate at $5$\,$\mu$m (pink) and $20$\,$\mu$m (green) from the graphene ribbon edge. Solid (dashed) lines shows PL measurements as the gate potential decreases (increases). Each sweep is normalised to the first data point at $V_G = 4$\,V. A settling time of $2$\,s with $50$ camera cycles per data point was used for this sweep. (d) Difference in normalised PL across the GFET where the plotted value is the difference between the integrated PL at low gate potential, $V_G = 4$ to $2$\,V (band labeled PL$_+$ in (c)), and the PL at high gate potential, $V_G = -14$ to $-16$\,V (band labeled PL$_-$ in (c)), normalised by the value of PL$_+$. The PL is averaged across both sweep directions. The coloured squares indicate the areas for which the full PL curves are shown in (c). A settling time of $0.5$\,s with $5$ camera cycles per data point was used for this sweep to minimise fringe artefacts from drifting optics. (e) Profile of the change in normalised PL running across the graphene ribbon parallel to the gate (blue dashed line in (d)). The edges of the graphene ribbon (tears) are indicated by the black (grey) dashed lines. All scale bars are $20$\,$\mu$m.}
		\label{Fig3}
	\end{center}
\end{figure}

This measurement was made on the same device in which current densities were mapped. The gate potential is swept across from $4$\,V to $-16$\,V and back, encompassing the neutrality point of the device, which is seen at approximately $-12$\,V [Fig. \ref{Fig3}(a)]. The device resistance is left to equilibrate at $V_G = 4$\,V  under laser illumination prior to sweeping. Comparing snapshots of the accumulated PL images at the extrema of the gate potential sweep [Fig. \ref{Fig3}(b)] demonstrates a clear quenching of the PL near the graphene ribbon under the gate at the lower gate potential. Looking at area normalised PL curves around the device, we see that the PL is constant  above $V_G = -8$\,V, then declines by up to $25$\,\% in regions close to, but not directly beneath the graphene [Fig.\ref{Fig3}(c)]. To map the extent of this effect, we take a normalised difference between the PL averaged across a $2$\,V window at either end of the sweep, PL$_-$ and PL$_+$ in Fig. \ref{Fig3}(c), and plot this difference across the full field of view [Fig. \ref{Fig3}(d)]. The normalised PL difference images shows the PL switching effect occurs only under the top gate, in regions un-screened by the graphene. Surprisingly this effect extends laterally from the graphene ribbon by up to $20$\,$\mu$m [Fig. \ref{Fig3}(e)]. We note that the magnitude of this PL change increases throughout the lifetime of the device (appendix \ref{eleccharac}).

\begin{figure}[h!]
	\begin{center}
		\includegraphics[width=1.0\columnwidth]{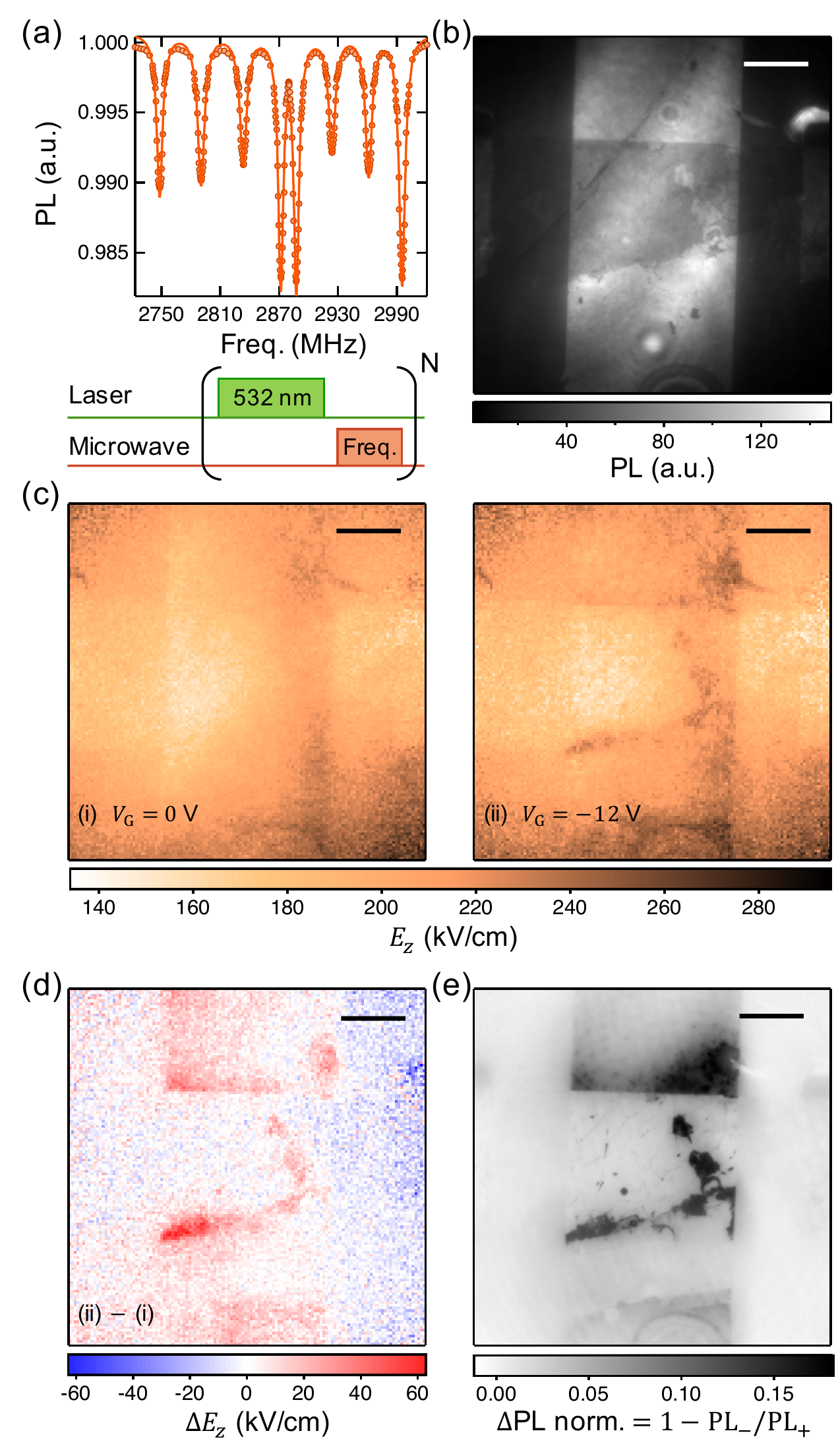}
		\caption{\textbf{Electric field measurements:} (a) ODMR spectrum optimised for electric field sensing by minimising the bias magnetic field projection along one NV-family and enhancing their optical contrast (upper). ODMR pulse sequence (lower). (b) PL image of the electric field mapped device taken under CW illumination. (c) Maps of the electric field in the z-direction at two applied gate potentials, (i) $V_G = 0$\,V, and (ii) $V_G = -12$\,V. The map was extracted from the ODMR data, presuming the field lay only in the z-direction for simplicity. A small source-drain potential was applied for each measurement $V_{SD} = 100$\,mV to track the conductivity of the device. (e) Map of the difference in electric fields measured at $V_G = -12$\,V and $V_G = 0$\,V. (f) Difference in normalised PL between the PL at low gate potential, $V_G = -1$ to $+1$\,V and the PL at high gate potential, $V_G = -11$ to $-13$\,V, normalised by the value at lower gate potentials, as in Fig. \ref{Fig3}. The data was taken using a $0.5$\,s settling time and $5$ camera accumulations per gate potential value. All scale bars are $20$\,$\mu$m.}
		\label{Fig4}
	\end{center}
\end{figure}

A likely explanation for this gate potential dependent PL, is that varying the gate potential affects the charge distribution at the diamond-oxide interface, and hence the degree of band bending across the NV-layer \cite{Broadway2018b}. This effect has been investigated previously by varying the surface chemistry of the diamond \cite{Hauf2011,Cui2013} and by electrostatic gating of NV centres \cite{Grotz2012,Hauf2014,Pfender2017}. Here, we suggest that an increasing gate potential populates an acceptor layer at the diamond surface \cite{Stacey2018a}, given the low carrier density within the implanted region, and hence increases the band bending across the NV-layer. As the band bending increases, the Fermi level falls below the NV$^-$ charge state at greater depths within the diamond. Once this occurs around the mean depth of the NV ensemble, the PL of the layer is reduced significantly as NV$^0$ becomes the dominant charge state. The comparative abundance of charge carriers in the graphene is expected to screen this effect, and hence we only observe the PL switching close to the gated device, but not directly under the graphene ribbon. We note that the charge state stability and dynamics under illumination and in the dark may also be altered by the gate potential through the local charge environment within the diamond \cite{Dhomkar2018a,Bluvstein2019}.

\section{Electric field measurements}

In order to test the outlined hypothesis, a direct measurement of the electric field was made. The electric field across the NV-layer can be measured in a direct and quantitative manner via the Stark effect on the NV spin-states \cite{Doherty2013}. To do this, we performed an ODMR measurement, as for the previous \O rsted field measurements, but with the biasing magnetic field oriented such that the magnetic field projection along a single NV-family is minimised, enhancing its electric-field sensitivity \cite{Dolde2011,Broadway2018b}, but still allowing each spin-state transition of that family to be resolved [Fig. \ref{Fig4}(a)]. The optical contrast of these transitions was selectively enhanced by rotating the linear polarisation of the excitation laser to further improve sensitivity \cite{Alegre2007}.

The electric field was mapped across a single GFET device [Fig. \ref{Fig4}(b)] for applied gate potentials of $V_G = 0$\,V, and $V_G = -12$\,V [Fig. \ref{Fig4}(c)]. In both measurements, the electric field is approximately $40$\,kV/cm less in the NV-layer under the graphene ribbon than that directly under the oxide. This may be partially due to the graphene modifying surface space-charge distribution at the diamond surface, and hence decreasing the band bending, however, it is more likely to be fully explained by FRET effect with the graphene quenching PL from NV centres closer to the surface, biasing the PL collection to deeper NVs which see a lower electric field. At $V_G = -12$\,V, the electric field increases at the edges and within tears in the graphene ribbon under the top gate. These features are highlighted in taking a subtraction of the two maps [Fig. \ref{Fig4}(d)], which shows an enhanced electric field up to $60$\,kV/cm in these regions, and essentially no change under the graphene ribbon. 

Comparing the electric field map to the difference in normalised PL of the same device [Fig. \ref{Fig4}(e)], we see a strong correlation in the location of features showing an enhanced electric field and reduced PL at high gate potential. The correspondence between the directly measured electric field and the gate potential dependent PL validates the claim that PL dependence on gate potential arises from electric field mediated band bending within the diamond \cite{Broadway2018b}. The spatial distribution of this effect, namely that it extends up to $20$\,$\mu$m from the gated device, is less trivial. A finite element method simulation of the purely dielectric response of the system indicates that the electric field from the gate should be less than $10$\,kV/cm within the NV-layer for distances larger than $5$\,$\mu$m from the graphene edge, and thus cannot account for the observations (appendix \ref{COMSOL}). For this reason, we suggest that the measured electric field originates from trapped charge at the diamond-oxide interface that accumulates above a threshold gate potential. Although the exact mechanism is unclear, this accumulation may result from charge diffusion (appendix \ref{eleccharac}) either through the oxide or the diamond itself, mediated by photo-excitation of the laser \cite{Jayakumar2016}.

\section{Outlook}

This work highlights the invasiveness of NV microscopy in the case of GFETs fabricated directly on an NV-diamond substrate. Namely, we found that control over the average doping in the graphene layer is strongly affected by a gate potential dependent photo-doping effect, while degradation in the device from measuring at high gate potentials over time scales required for NV imaging limits the ability to probe a single device in numerous scenarios. Additionally, we see evidence of a complex electrostatic response at the oxide-graphene and diamond-oxide interfaces that is not limited to ALD Al$_2$O$_3$ dielectric \cite{Kim2013,Tiberj2013,Ju2014a}, which raises the possibility of uncontrolled spatially-dependent doping variations. For future applications of NV microscopy where a precise, reliable control over the doping is required, these effects must be mitigated. One possible solution is to decouple the GFET from the diamond by capping the diamond with a metal-oxide bilayer before fabricating the GFET. The extra metallic layer would prevent laser radiation from reaching the GFET \cite{Barry2017}, drastically reduce any charge transfer effect at the diamond surface by providing an electron reservoir, and could even serve as a bottom gate for the GFET. A downside of this solution is that the graphene layer can no longer be visualised optically through the FRET effect. Further enhancements can made by encapsulating the graphene ribbon in hexagonal boron nitride (h-BN), which allows the h-BN to be used as a gate dielectric, and is known to improve graphene quality \cite{Andersen2019,Ku2019}.

A remaining issue arises from the large source-drain currents typically required for NV magnetometry. This requirement could be relaxed by improving the sensitivity of the NV sensing layer. In the present work, we aimed for a mean graphene-NV distance of only $10-20$\,nm in order to preserve a sizable FRET effect facilitating imaging, but this requirement has a direct impact on sensitivity by limiting the maximum number of NV centres without compromising the NV spin coherence \cite{Tetienne2018}. However, in principle thicker NV layers (e.g. $200$\,nm) can be employed without deteriorating the spatial resolution which would remain limited by diffraction ($\approx300$nm) \cite{Tetienne2018b}. For instance, the optimised NV-layer in Ref. \cite{Kleinsasser2016} would provide a $10$-fold improvement in magnetic sensitivity. With a further increase in collected PL signal due to the extra metallic layer, NV measurements with source-drain currents in the $\mu$A range can be envisaged. Implementing these solutions may allow for minimally-invasive wide-field NV microscopy of GFETs and other electrical devices based on two-dimensional materials.

We thank Daniel J. McCloskey and Alastair Stacey for useful discussions. This work was supported by the Australian Research Council (ARC) through Grants No. DE170100129, No. CE170100012, and No. FL130100119. This work was performed in part at the Melbourne Centre for Nanofabrication (MCN) in the Victorian Node of the Australian National Fabrication Facility (ANFF). D.A.B. and S.E.L. are supported by an Australian Government Research Training Program Scholarship.

\bibliographystyle{apsrev4-1}
\bibliography{GrapheneFETpaper}
	   
\appendix
	   
\section{Diamond samples}\label{sample}

The NV-diamond samples used in these experiments were made from $4$\,mm $\times$ $4$\,mm $\times$ $50$\,$\mu$m electronic-grade ([N]~$<1$~ppb) single-crystal diamond plates with \{$110$\} edges and a ($100$) top facet, purchased from Delaware Diamond Knives. The diamond surfaces were polished to a roughness $<2$\,nm \cite{Lillie2018}. The plates were laser cut into smaller $2$\,mm $\times$ $2$\,mm $\times$ $50$\,$\mu$m plates, acid cleaned ($15$ minutes in a boiling mixture of sulphuric acid and sodium nitrate), and implanted with $^{15}$N$^+$ ions (InnovIon) at an energy of $6$\,keV and a fluence of $10^{13}$~ions/cm$^2$ with a tilt angle of $7^\circ$. Such energy corresponds to a mean implantation depth of $10$-$20$~nm~\cite{Tetienne2018}. Following implantation, the diamonds were annealed in a vacuum of $\sim10^{-5}$~Torr to form the NV centres, using the following sequence: $6$\,h at $400^\circ$C, $2$\,h ramp to $800^\circ$C, $6$\,h at $800^\circ$C, $2$\,h ramp to $1100^\circ$C, $2$\,h at $1100^\circ$C, $2$\,h ramp to room temperature. To remove the graphitic layer formed during the annealing at the elevated temperatures, the samples were acid cleaned (as before).

\section{Fabrication}\label{FETfab}

The GFETs in this work all consisted of monolayer polycrystaline graphene ribbons, Cr/Au source drain contacts, and an $80$\,nm Al$_2$O$_3$ gate oxide with a Cr/Au top gate contact. Two sets of $10$  GFETs were fabricated on two different NV-diamond substrates, labeled as diamond \#$200$ and diamond \#$230$. On diamond \#$200$ the graphene ribbons were $50$\,$\mu$m $\times$ $500$\,$\mu$m, with the source contacts evaporated on top of the graphene. On diamond \#$230$ the ribbons were $20$\,$\mu$m $\times$ $500$\,$\mu$m, and the graphene was transferred on top of already existing contacts.

Graphene was transferred from commercially available (Graphenea) mono-layer graphene grown by chemical vapour deposition (CVD) on a copper foil using a standard wet chemical technique \cite{Liang2011}. Prior to transfer the graphene was spin-coated with a PMMA A$4$ ($950$) protective layer. The copper foil was etched in a $0.5$\% wt. Fe(NO$_3$)$_3$ solution for $24$\,hours. The sample was transferred (with a clean Si wafer) through multiple DI rinses, a dilute RCA2 ($34$\%HCl:H$_2$O$_2$:DI $1$:$1$:$5$) cleaning step and further rinsing before being transferred to a $2$\,mm $\times$ $2$\,mm NV-diamond substrate, mounted on a $300$\,nm SiO$_2$/Si wafer. The sample was left to dry over $48$\,hours.

The Al$_2$O$_3$ top gate oxide was grown via atomic layer deposition (ALD), with TMA and water precursors at $200^\circ$C. Nucleation issues on the graphene were mitigated by increasing the water residence time with a double pulse within the first $50$ cycles \cite{Aria2016}. The bonding pads were exposed by etching the oxide in a $8$\% NH$_4$OH solution at $50^\circ$C for $25$\,min.

Contacts, bonding pads, and the top gate were fabricated with photolithography using TI$35$E photoresist, followed by thermal electron-beam evaporation and liftoff of Cr/Au ($10$/$70$\,nm). The graphene was also patterned with photolithgraphy, using SU$8$ (negative tone photoresist) on a protective PMMA layer to create a removable hard mask for etching in an oxygen plasma asher ($750$\,W, $10$\,sccm O$_2$ in Ar).


\section{Electrical characterisation and photo-doping}\label{eleccharac}

\begin{figure*}
	\begin{center}
		\includegraphics[width=1.0\textwidth]{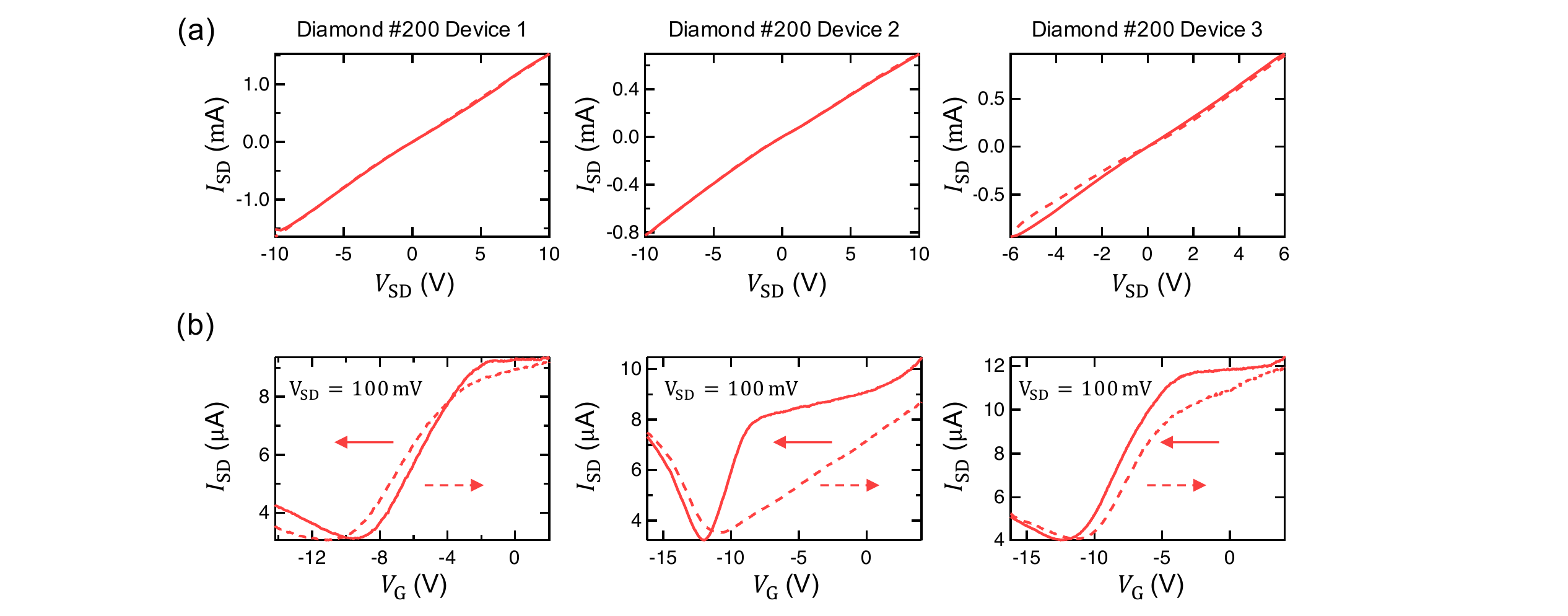}
		\caption{\textbf{Electrical characterisation of GFETs:} (a) $I_{SD}$ versus $V_{SD}$ measurements of three devices from diamond \#$200$. Averaging linear fits of the increasing and decreasing sweeps gives resistances of $6.38$\,k$\Omega$, $13.21$\,k$\Omega$, and $6.49$\,k$\Omega$ for devices $1$, $2$, and $3$ respectively. Solid (dashed) lines indicate increasing (decreasing) $V_{SD}$ sweeps. (b) $I_{SD}$ versus $V_G$ sweeps showing neutrality points at $V_G \approx -11$\,V for the same three devices respectively, measured with a small source-drain current, $I_{SD} = 100$\,mV under CW laser illumination. Solid (dashed) lines indicate decreasing (increasing) $V_G$ sweeps. The device current was left to equilibrate at the initial point of each sweep prior to sweeping the gate potential.} 
		\label{FigS1}
	\end{center}
\end{figure*}

The GFETs were characterised electrically throughout their lifetime to track their resistance and doping. $I_{SD}$ versus $V_{SD}$ curves were measured for all devices prior to imaging, showing reasonable Ohmic behavior of the graphene ribbons [Fig. \ref{FigS1}(a)]. Linear fits of the $I_{SD}$ versus $V_{SD}$ curves give resistances of $6.38$\,k$\Omega$, $13.21$\,k$\Omega$, $6.49$\,k$\Omega$ for devices $1$, $2$, and $3$ respectively on diamond \#$200$. We note that device resistances typically increased throughout measurement, particularly after measuring at high source drain currents ($>500$\,$\mu$A) and high gate potentials ($|V_G|>8$\,V), which was observed to cause tearing in the ribbon (appendix \ref{currentdensity}). Conductivity minima were also found for these devices prior to imaging, by measuring $I_{SD}$ as a function of $V_G$ for a small source-drain potential, $V_{SD} = 100$\,mV, under CW laser illumination [Fig. \ref{FigS1}(b)]. Devices $1$, $2$, and $3$ on diamond \#$200$ show conductivity minima at $V_G \approx -11$\,V.

\begin{figure*}
	\begin{center}
		\includegraphics[width=1.0\textwidth]{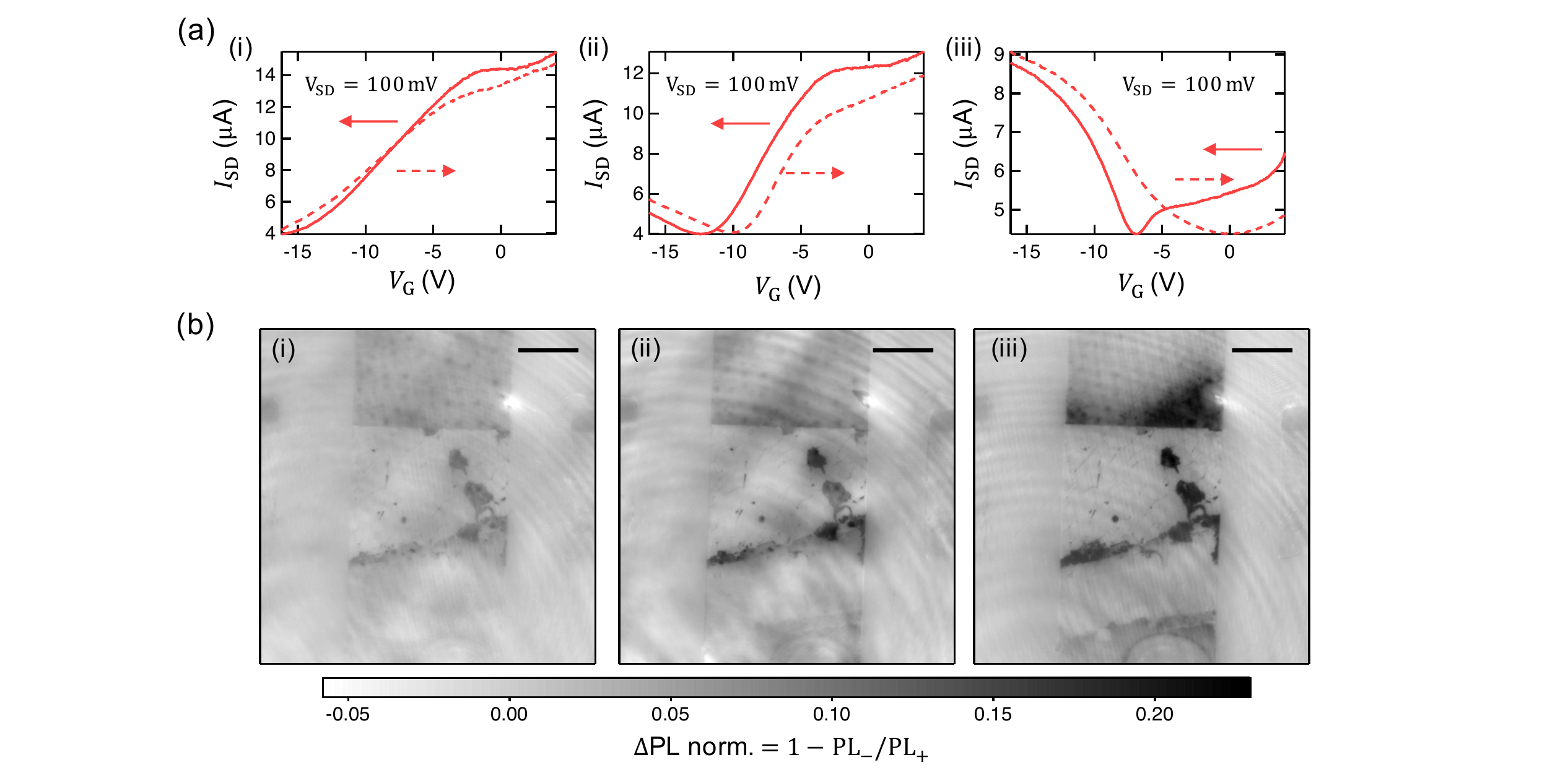}
		\caption{\textbf{Evolution of neutrality point and gate potential dependent PL:} (a) $I_{SD}$ versus $V_G$ curves for device $3$ on diamond \#$200$: (i) prior to any sustained measurements; (ii) after electric field mapping at $V_G = 0$\,V and $V_G = -8$\,V; and (iii) after electric field mapping at $V_G = -12$\,V. All measurements were taken under CW laser illumination with $V_{SD} = 100$\,mV. (b) (i)-(iii) Difference in normalised PL maps of device $3$ on diamond \#$200$ taken immediately after the measurements in (a). The plotted value is the difference between the integrated PL at high gate potential, $V_G = 2$ to $4$\,V (PL$_+$) and the PL at low gate potential, $V_G = -14$ to $-16$\,V (PL$_-$), normalised by the value at higher gate potential. The PL is averaged across both sweep directions. A settling time of $2.0$\,s with $50$ camera cycles per data point was used for this sweep. All scale bars are $20$\,$\mu$m.}
		\label{FigS2}
	\end{center}
\end{figure*}

Throughout the course of imaging a device, $I_{SD}$ versus $V_G$ curves were measured regularly under CW laser illumination to track changes in the effective doping of the device under the relevant imaging conditions. We observe that sustained imaging of the devices, which requires a prolonged exposure to some combination of laser, gate potential, and source drain current, resulted in shifts of the conductivity minima, in addition to an enhanced hysteresis in the $I_{SD}$ versus $V_G$ curves measured under laser illumination [Fig. \ref{FigS2}(a)]. In conjunction with this effect, we observe an enhanced quenching of the NV PL at low gate potentials over the same time frame [Fig. \ref{FigS2}(b)]. The difference in normalised PL maps shown are produced in the same fashion as those presented in Fig. \ref{Fig3} in the main text, and demonstrate an increase in the magnitude of the PL change, and its lateral extent, after prolonged imaging. Importantly, we note that the onset of this effect is occurs consistently at a threshold gate potential of $-8$\,V.

An explanation for these changes to the effective device doping and enhanced hysteresis is that there is a photon assisted charge transfer between the graphene and oxide, similar to the optical-doping seen with other gate dielectrics and substrates \cite{Kim2013,Tiberj2013,Ju2014a}, which has some dependence on the applied gate potential at the time of illumination. To test this, a single device was exposed to CW laser illumination under a fixed gate potential ($V_G^{\textrm{PD}}$) and left to equilibrate over a $6$\,min time period, while a small source-drain potential, $V_{SD} = 100$\,mV, was applied to measure the current through the device. The laser was then turned off, and then the gate potential swept between $-10$ and $+10$\,V, while measuring the source-drain current to identify the conductivity minimum ($V_G^{\textrm{min}}$). The measurement was repeated for each photo-doping gate potential to measure the transport curve by sweeping the gate potential in the opposite direction, which was seen to systematically shift the location of the minima. We attribute this to the presence of stray light incident on the sample unpinning the photo-doping during the sweep, and additional trapped charge at the interfaces. The sample was photo-doped at $V_G^{\textrm{PD}} = 0$\,V between each measurement, such that subsequent photo-doping proceeded from similar initial conditions.

\begin{figure}
	\begin{center}
		\includegraphics[width=1.0\columnwidth]{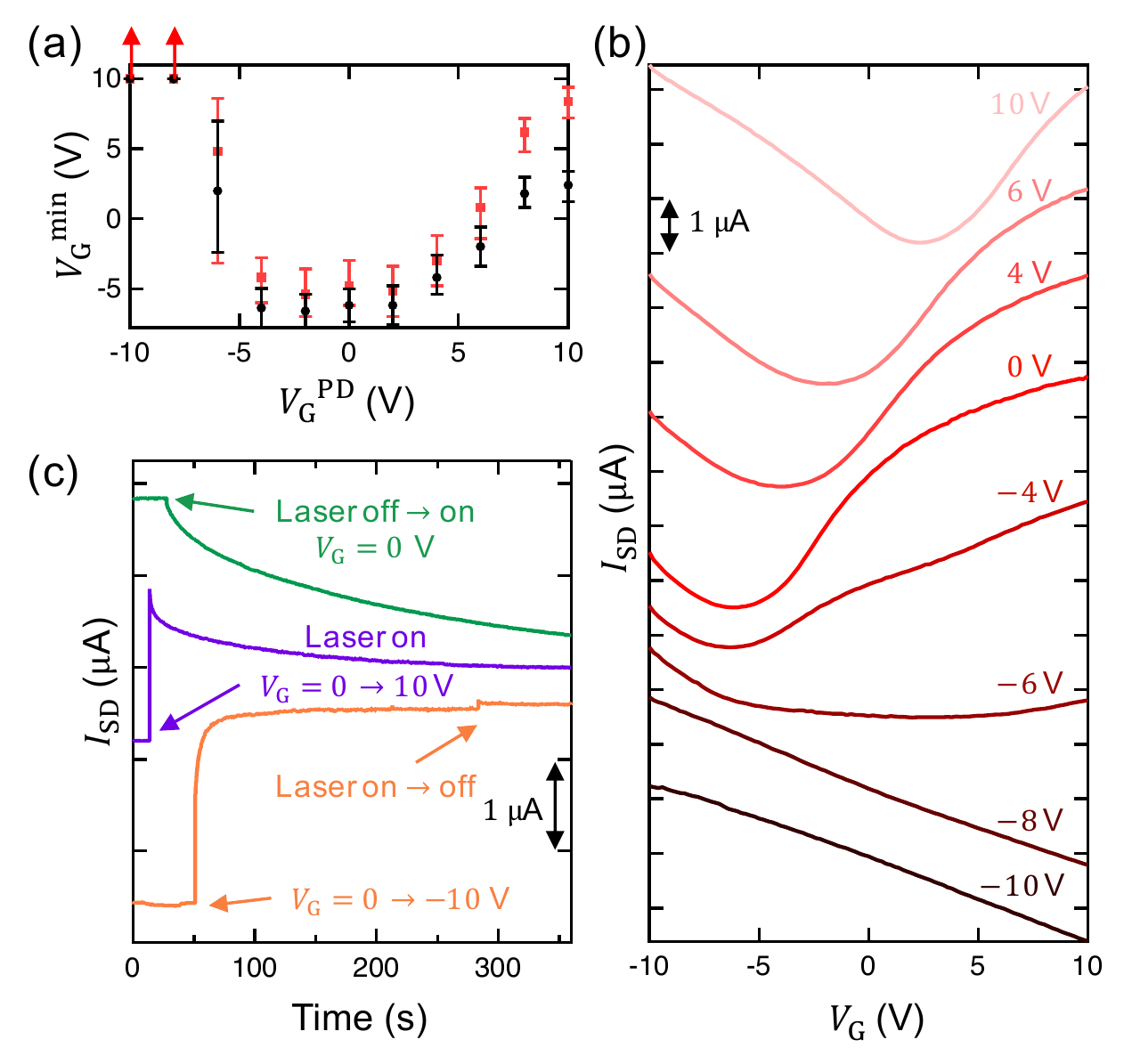}
		\caption{\textbf{Photo-doping of a single GFET device:} (a) Measured gate potential of the conductivity minimum, $V_G^{\textrm{min}}$, as a function of the gate potential applied during photo-doping, $V_G^{\textrm{PD}}$. $V_G^{\textrm{min}}$ are given as measured from an increasing (black) and decreasing (red) $V_G$ sweeps. The error bars show the width of the transport curve at $+0.1$\,$\mu$A from the conductivity minimum. (b). Individual $I_{SD}$ versus $V_G$ curves for a single GFET device as measured in the dark following photo-doping at the labeled gate potential, $V_G^{\textrm{PD}}$. The curves were all measured with increasing gate potential sweeps, and correspond to the red data set shown in (a). The curves are offset for clarity, and the minimum $I_{SD}$ for each measurement are $2.69$, $3.10$, $3.22$, $3.01$, $3.78$, $4.49$, $3.79$, and $4.39$\,$\mu$A in descending order of the photo-doping gate potentials. (c) Time traces of $I_{SD}$ show effect of photo-doping parameters. $V_G = 0$\,V for the entirety of the green curve, and the laser is turned on at the indicated time point, leading to a bi-exponential decay in the device current. The CW laser is on initially for the purple (orange) curves, and the gate potential is switched from $V_G = 0$\,V to $+10$\,V ($-10$\,V). The laser is turned off in the orange time trace at approximately $280$\,s. The minimum current in each time trace is $3.20$, $3.39$, and $2.81$\,$\mu$A for the green, purple and orange curves respectively.}
		\label{FigS3}
	\end{center}
\end{figure}

The conductivity minima were extracted from transport measurements for $V_G^{\textrm{PD}}$ between $-10$ and $10$\,V [Fig. \ref{FigS3}(a)]. Photo-doping at $V_G^{\textrm{PD}} = 0$\,V gives an n-type doped graphene channel with conductivity minimum at $V_G^{\textrm{min}} \approx -5$\,V. The minimum shifts to higher gate potentials as $V_G^{\textrm{PD}}$ is increased, giving a p-type device for $V_G^{\textrm{PD}} \geq 8$. At negative $V_G^{\textrm{PD}}$, we identify a strong threshold at $V_G^{\textrm{PD}} = -6$\,V where conductivity minimum shifts to positive gate potentials, $V_G^{\textrm{min}} > 10$\,V, a regime we are unable to measure due to large leakage currents through the gate oxide. We note that this threshold closely corresponds to the onset of the PL switching effect, and hence suggest that both phenomena may be due an accumulation of holes within the oxide and at the diamond-oxide and graphene-oxide interfaces. The full transport curves measured with increasing gate potential sweeps following different photo-doping gate potentials highlight this effect [Fig. \ref{FigS3}(b)], where the conductivity minima are beyond the range of the sweep for photo-doping $V_G^{\textrm{PD}} \leq -8$\,V.

The time scales over which the photo-doping of the graphene channel occurred were measured by tracking the source-drain current within the $6$\,min window during which the CW laser was switched on or off, and the gate potential set [Fig. \ref{FigS3}(c)]. The time evolution of $I_{SD}$ at $V_G = 0$\,V (green curve) demonstrates the necessity of the laser in the photo-doping effect, which initiates a decrease in conductivity of the device over minute long time-scales from some initially pinned value. The gate potential dependence is evident in the time-trace when the gate potential is set from $V_G = 0$\,V to $10$\,V (purple curve) and $-10$\,V (orange curve) under CW laser illumination, after having equilibrated under laser illumination at $V_G^{\textrm{PD}} = 0$\,V. As the gate potential is changed there is a large change in conductivity, as the system jumps to a doping level dictated by the $I_{SD}$ versus $V_G$ curve photo-doped at $V_G^{\textrm{PD}} = 0$\,V, and then a slower evolution of the conductivity, which in each case corresponds to a shifting of the conductivity minimum to higher gate potentials and hence reducing (increasing) the $I_{SD}$ in the $V_G = 10$\,V ($-10$\,V) case. The time evolution of $I_{SD}$ is best fit by a bi-exponential with a fast and slow component acting on time scales of $10$s and $100$s respectively, the exact values of which are sensitive to the initial conditions of the photo-doping. We also observe a jump in the conductivity of the device when the laser is turned off after the device has equilibrated under CW illumination (seen at $280$\,s, orange curve), suggesting a gate dielectric dynamic faster than our time-resolution ($0.08$\,s), however, the system equilibrates to a similar doping level as reached under illumination. For this reason, we conclude that the doping achieved by a set gate potential under CW laser illumination is maintained when the illumination is then pulsed as required for NV measurements such as ODMR. Time-traces of the device source-drain current throughout ODMR measurements endorse this.

\section{Current density reconstruction}\label{currentdensity}

The current density maps presented in the main text were produced using a method established in previous work \cite{Tetienne2017,Tetienne2019}, which proceeds in the following manner. The ODMR spectrum at each pixel was fit with an eight Lorentzian sum, and the frequencies extracted. The magnetic field projection along the four NV-family orientations was calculated from the Zeeman splitting of each frequency pair from the zero field resonance at $2870$\,MHz. The field was then converted to Cartesian coordinates using the three most split NV-families, having previously determined their orientation relative to the diamond surface by measuring a field of known orientation \cite{Steinert2010,Chipaux2015}.

To reconstruct the current density from the measured magnetic field, we invert the Biot-Savart law in Fourier space \cite{Roth1989,Nowodzinski2015}. Here, we take only the $B_z$ component of the measured magnetic field and linearly extrapolate the remnant field in the y-direction in order to minimise truncation artefacts in the Fourier transform \cite{Tetienne2019}. The Fourier space current densities in the x- and y-directions are calculated trivially from the transformed $B_z$, and their inverse Fourier transform gives us the real space densities \cite{Tetienne2019}. 

Previous work has highlighted an apparent delocalisation of current from metallic systems on the diamond surface to the diamond itself, as measured by this same technique \cite{Tetienne2019}. Here, we note that all current densities plotted in this work represents the total reconstructed current density (above and below the NV-layer), but most of the current was found to lie above the measuring NV-layer. This scenario was consistent as the gate potential was varied. 

Each current density map produced arose from two separate ODMR measurements of the device at the given gate potential; one with and the other without the injected source-drain current. This was done to control for any magnet or electric field features not associated with the carrier transport in the GFET. The subtraction of the background measurement from the signal was performed prior to converting the NV-family field projections to Cartesian coordinates.

\begin{figure*}
	\begin{center}
		\includegraphics[width=1.0\textwidth]{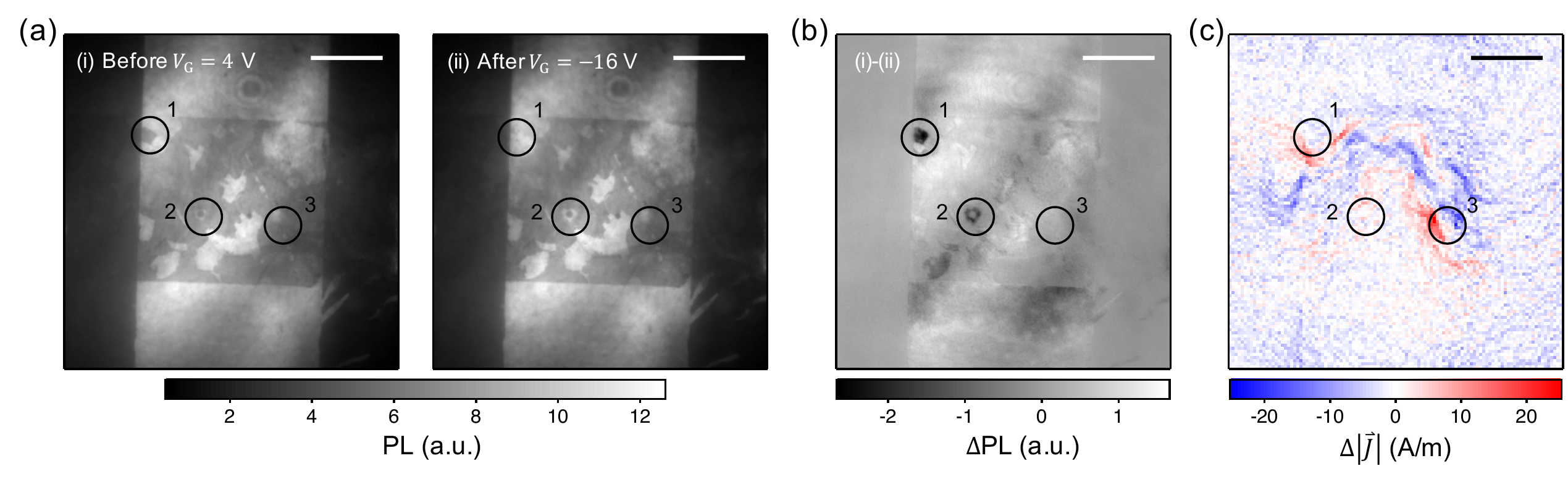}
		\caption{\textbf{PL imaging throughout current density mapping:} (a) PL images of the current density mapped GFET device at $V_G = 0$\,V taken: (i) before mapping at $I_{SD} = 700$\,$\mu$A and $V_G = +4$\,V; and (ii) after mapping at $I_{SD} = 700$\,$\mu$A and $V_G = -16$\,V. (b) Subtraction of (i) and (ii). (c) Difference in norm current densities under doping conditions $V_G = +4$\,V and $V_G = -16$\,V for comparison with the subtracted PL map. All scale bars are $20$\,$\mu$m.}
		\label{FigS4}
	\end{center}
\end{figure*}

The current density maps produced by this method showed differing density distributions at each doping condition [Fig. \ref{Fig2}]. A simple explanation for these differences is that they are due to degradation of the graphene throughout the measurement, which was often observed following sustained high source-drain currents ($>500$\,$\mu$A) and at high gate potentials ($\abs{V_G}>8$\,V). To determine whether the observed differences arise from significant tearing of the graphene or not, we compare PL images of the device taken under the same conditions prior to mapping at $V_G = +4$\,V, and after mapping at $V_G = -16$\,V [Fig. \ref{FigS4}(a)]. A subtraction of these two PL images highlights two regions (marked $1$ and $2$) in the gated section of the graphene ribbon where the graphene is no longer visible via FRET interaction with the NV-layer in the later measurement [Fig. \ref{FigS4}(b)]. PL imaging between the two current measurements indicate that these changes occurred during the $V_G = -16$\,V measurement. The fringes visible across the device in the subtraction are due to a slight shift in the optics between measurements. 

Comparing the PL subtraction to the current density subtraction [Fig. \ref{FigS4}(c)] shows that there is a deviation in current path close to one of the tears ($1$). Interestingly, the region showing the most distinct change in current path (marked $3$) shows very little PL change, indicating that it is not associated with a degradation of the graphene visible at our imaging resolution. Further investigation into the cause of this current deviation is made difficult by the gradual deterioration of the device throughout measurement, which precludes repeat measurements, and motivates a new generation of devices which better isolate the graphene from the diamond substrate and oxide-interface. 

\section{Electric field simulations}\label{COMSOL}

\begin{figure}
	\begin{center}
		\includegraphics[width=1.0\columnwidth]{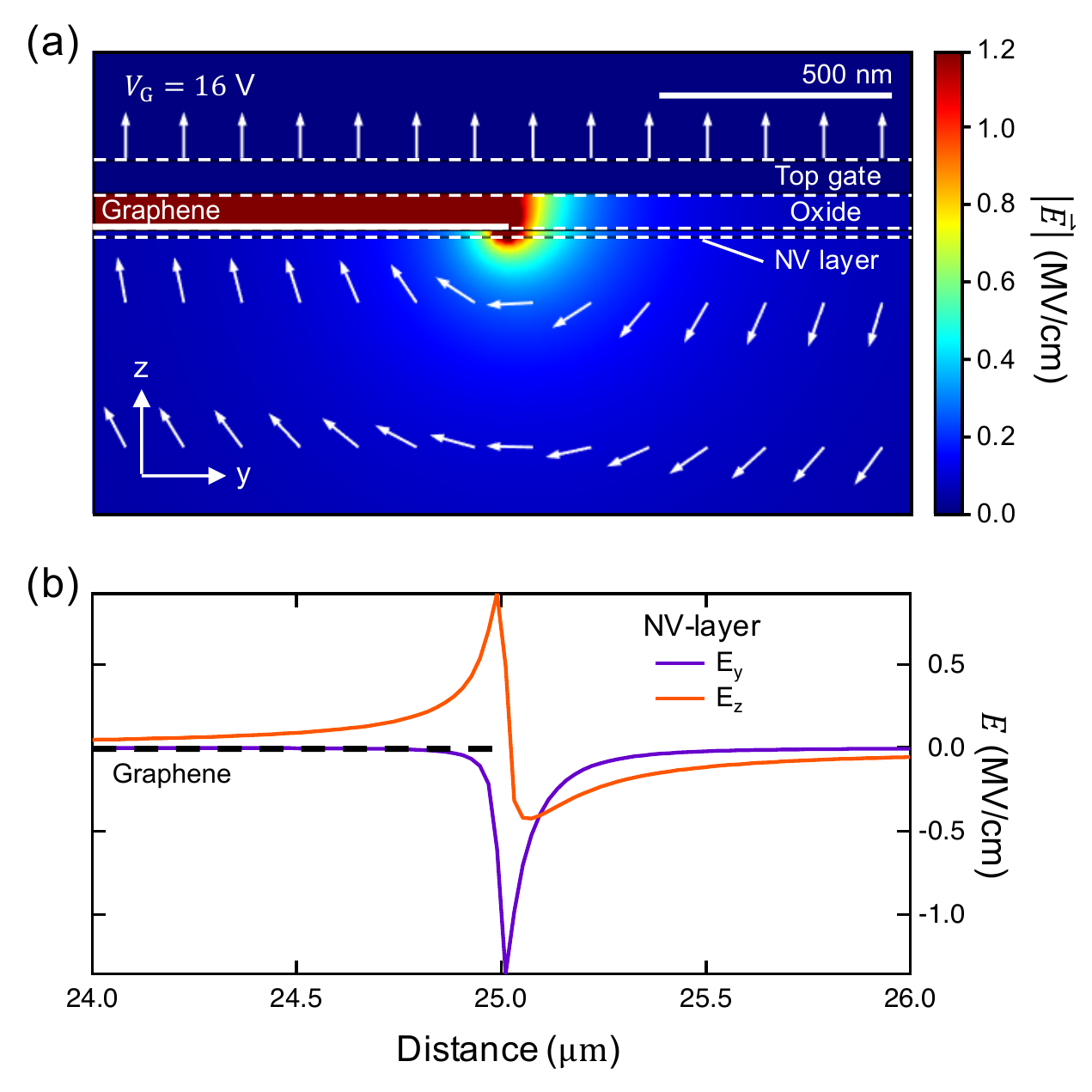}
		\caption{\textbf{COMSOL simulation of top gate electric field:} (a) Simulated electric field magnitude plot for $V_G = +16$\,V within a plane running perpendicular to the graphene ribbon edge, in the middle of the top gate (i.e. corresponding to a vertical line cut of the images shown in the main text). The graphene extends from the left-hand-side to the middle of the plot, whereas the top gate, oxide, and diamond cover the full region. The arrows show the projection of the electric field in the plane. The aspect ratio is $1:1$, and the colour scale has been capped at $1.2$\,MV/cm to highlight features at lower field magnitudes. (b) Electric field components in the z- and y-directions in the NV-layer ($15$\,nm below the diamond surface) at $V_G = +16$\,V across the width of the image shown in (a). The graphene ribbon edge lies at $25$\,$\mu$m.}
		\label{FigS5}
	\end{center}
\end{figure}

To determine the spatial distribution of the electric field from the top gate, such that it can be compared to that measured by the NV-layer, a finite element method (FEM) simulation was performed using COMSOL. The device geometry was replicated, where a $50$\,$\mu$m wide metallic top gate was separated from a graphene ribbon by $80$\,nm of Al$_2$O$_3$, all hosted on top of a $50$\,$\mu$m thick diamond substrate. The electric field distribution was calculated in a shell surrounding the device, using a high density mesh in the region of interest between the metallic planes and beneath the graphene plane, across the region containing the NV-layer.

The simulated electric field shows a high field strength between the parallel plate capacitor ($-2$\,MV/cm in the centre of oxide), which is screened from the diamond by the graphene plate [Fig. \ref{FigS5}(a)]. Appreciable field strengths exist in the diamond only at the edge of the graphene ribbon under the top gate, where the magnitude reaches $-0.9$\,MV/cm in the z-direction, but reverses in sign within a $300$\,nm length scale across the graphene ribbon edge [Fig. \ref{FigS5}(b)]. The component perpendicular to the graphene ribbon reaches $-1.2$\,MV/cm at the edge, but is laterally confined to $200$\,nm. Given the optical resolution limited imaging via the NV-layer PL, we do not expect to be able to resolve these features, and hence conclude that the NV-layer measurements should not be sensitive to electric field from the gate directly. Therefore, we propose that the enhanced electric field we measured, which correlates strongly with the gate potential dependent PL effect, must result from a change in the surface charge distribution at the diamond-oxide interface. 

\end{document}